\documentclass{WileyMSP-template}

\usepackage[acronym,nomain]{glossaries}
\setacronymstyle{long-short}
\newacronym{ac:ccd}{CCD}{Charge-Coupled Device}
\newacronym{ac:diw}{DIW}{Deionized Water}
\newacronym{ac:lod}{LOD}{Limit-of-Detection}
\newacronym{ac:pcf}{HC-PCF}{Hollow-Core Photonic Crystal Fiber}
\newacronym{ac:pcs}{PCS}{Photon Correlation Spectroscopy}
\newacronym{ac:pdms}{PDMS}{Polydimethylsiloxane}
\newacronym{ac:ps}{PS}{Polystyrene}
\newacronym{ac:sem}{SEM}{Scanning Electron Microscope}
\newacronym{ac:sers}{SERS}{Surface-Enhanced Raman Spectroscopy}

\begin{document}

\pagestyle{fancy}
\rhead{\includegraphics[width=2.5cm]{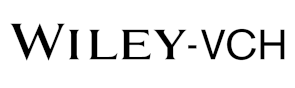}}

\title{Label-Free Spontaneous Raman Sensing in Photonic Crystal Fibers with Nanomolar Sensitivity}
\maketitle

% Author: Please give full first and last names for authors and include * after the name of all corresponding authors
\author{Basil G. Eleftheriades*}
\author{Emily E. Storey}
\author{Amr S. Helmy*}

% Affiliations: Please provide adacemic titles (Prof. or Dr.) for all authors where applicable, and include an institutional email address for all corresponding authors
\begin{affiliations}
Department of Electrical and Computer Engineering, University of Toronto, Toronto, Canada\\
Email Address: basil.eleftheriades@mail.utoronto.ca, a.helmy@utoronto.ca
\end{affiliations}

% Keywords: Please provide a minimum of three and a maximum of seven keywords, separated by commas

% Abstract should be written in the present tense and impersonal style (i.e., avoid we), and be at most 200 words long
\begin{abstract}

An approach to significantly enhance spontaneous Raman sensitivity through the formation of a thin film via thermophoresis along with evaporation at the facet of a Hollow-Core Photonic Crystal Fiber is reported for the first time. Sensitivity of detection is increased by more than 6 orders of magnitude for both organic and inorganic nanoparticles, facilitating the search for trace analytes in solution. Detection of two nanoparticles, Alumina and Polystyrene, is demonstrated down to 392 nM without the use of Surface-Enhanced Raman Spectroscopy or other chemical-based procedures. This new thin-film deposition approach simplifies the simultaneous detection and analysis of small trace compounds, a previously arduous task using conventional spontaneous Raman.

\end{abstract}

% Text: Please use section headings and subheadings as specified below. For communications, all section headings apart from Experimental Section should be removed
% Please make the first reference to a display item bold: \textbf{Figure 1}
% Do not abbreviate Figure, Equation, etc.; display items are always singular, i.e., Figure 1 and 2.
% Equations are always singular, i.e., Equation 1 and 2, and should be inserted using the {equation} environment, not as graphics
% Please do not use footnotes in the text, additional information can be added to the Reference list.

\section{Introduction} % word count 462
	Nanoparticles constitute a pivotal building block in existing and emerging structured material systems. Their utility spans the entire spectrum of applications from health to industrial and agriculture domains \cite{Cabeza2015,Burklew2012}. Monitoring nanoparticles, be it during synthesis, utilization, or as a contaminant, has emerged as a pressing necessity given their wide-spread deployment \cite{Hood2004}. This need and ability to carry out in-situ monitoring is in contrast to techniques which can characterize a stand alone sample of nanoparticles. Numerous spectroscopy and microscopy techniques can be utilized for such tasks  
	
	A plethora of techniques have been established to analyse a sample nanoparticles, but limitations prevent facile comprehensive characterization. One of the most prevalent techniques is X-ray diffraction, which provides information on nanoparticle size, composition, and crystalline structure \cite{Mourdikoudis2018}. Particles smaller than 3 nm or larger than 50 nm are incompatible with X-ray diffraction. Moreover, it is not suited to trace nanoparticle detection or concentration determination. Other characterization methods require similar compromise; several techniques are necessary when a certain combination of parameters is needed and are therefore not suitable for in-situ monitoring. 
	
	Optical techniques are useful for nanoparticle monitoring tasks. For example photon correlation Fourier spectroscopy has been proven as a useful tool to extract single nanoparticle emission linewidths  \cite{Vandenabeele2007}. In addition, polarized Photoluminescence has also been useful in studying the order and orientation of nanoparticles \cite{DeBeer2011}. Another approach utilizes Spontaneous Raman in, still and circulating fluid, liquid core waveguides and has enabled studies of nanomaterials with great sensitivity compared to conventional Raman \cite{Evans2008, Tang2018}.
	
	Raman spectroscopy is a versatile characterization technique, non-destructively providing information on the molecular structure of materials with exceptional specificity \cite{Gouadec2007}. Raman (inelastic) scattering, however, occurs at least 10 orders of magnitude less frequently than Rayleigh (elastic) scattering. The consequence is a faint scattering signal, which is particularly exacerbated in fluidic samples due to lower molecular density.  As such, Raman is primarily associated with solid samples. Raman's potential in liquid and gaseous samples has yet to be fully realized: signal enhancement techniques such as \gls{ac:sers} require extensive substrate or sample preparation and lack repeatability, hindering reliable characterization \cite{Fan2020}. We propose here that the pairing of spontaneous Raman with optofluidics provides the necessary enhancement, offering a means to repeatably and quantitatively identify trace unknown compounds which may pose a substantial threat \cite{Lafuente2020}.
	
	Incorporating optofluidics with spontaneous Raman increases sensitivity for the full spectrum of Raman modes rather than a selection, without altering the native state of the analyte \cite{Mak2013}. Both the laser light and the liquid of interest are confined to the same waveguiding cavity, enabling a strong and efficient process of light-matter interaction. A key challenge with spontaneous optofluidic Raman is a limit of sensitivity, on the order of millimolar (mM) \cite{Mak2011}. 
	
\begin{figure}[t]
		\centering
		%This is a screengrab, I am missing the original from Basil. 
		\includegraphics[width=0.5\linewidth]{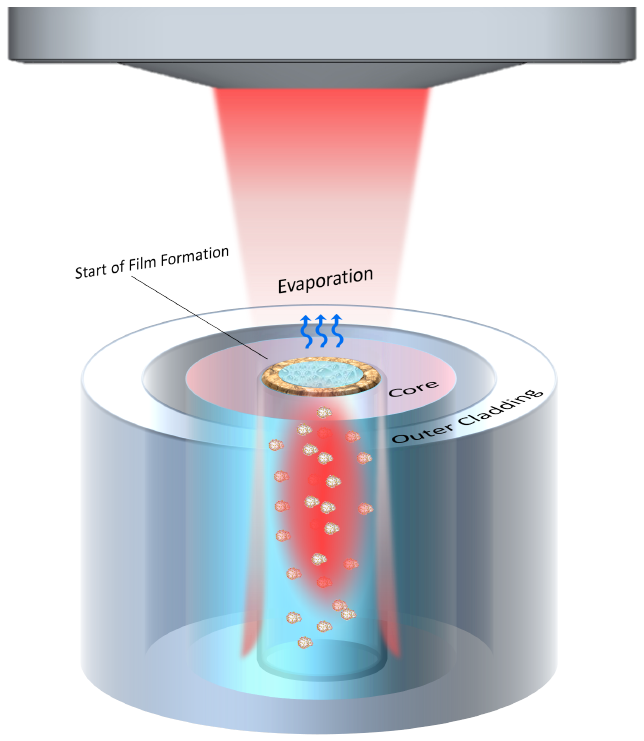} 
		\caption{Experimental setup: laser-induced thermophoresis atop a \acrfull{ac:pcf} prompts the formation of a thin film in the fiber core.}
		\label{fig:setup}
	\end{figure}

	In this work we demonstrate an approach utilizing thermophoresis along with evaporation and capillary effects to extend the sensitivity of spontaneous Raman in fluidic samples to the nanomolar (nM) range for nanoparticle sensing. The technique promotes assembly of a thin-film of nanoparticles, by means of the laser source which is used to induce scattering. The coffee-ring effect acts to aggregate the solute atop a \gls{ac:pcf}, and is depicted in \textbf{Figure \ref{fig:setup}}. The method is shown to be effective for a wide variety of samples both organic and inorganic. 
	
\section{Spontaneous Raman in Liquid Core Optical Fibers} % word count 60 + 520 = 580
	
	To assess the performance of the proposed technique, we benchmark the method against standard optofluidic \gls{ac:pcf} enhancement, in which the fluid and laser are co-located in the core of a \gls{ac:pcf} without further modification \cite{Mak2013}. The proposed technique will be evaluated in the following section, where we rely on targeted delivery of energy to promote the formation of a thin-film, rather than probing the sample along the entire length of a \gls{ac:pcf}. Samples used in this work shall be described, followed by their performance and lower limit of detection using the aforementioned \gls{ac:pcf} method. 

	\begin{figure*}[t]
		\centering
		\includegraphics[width=0.9\linewidth]{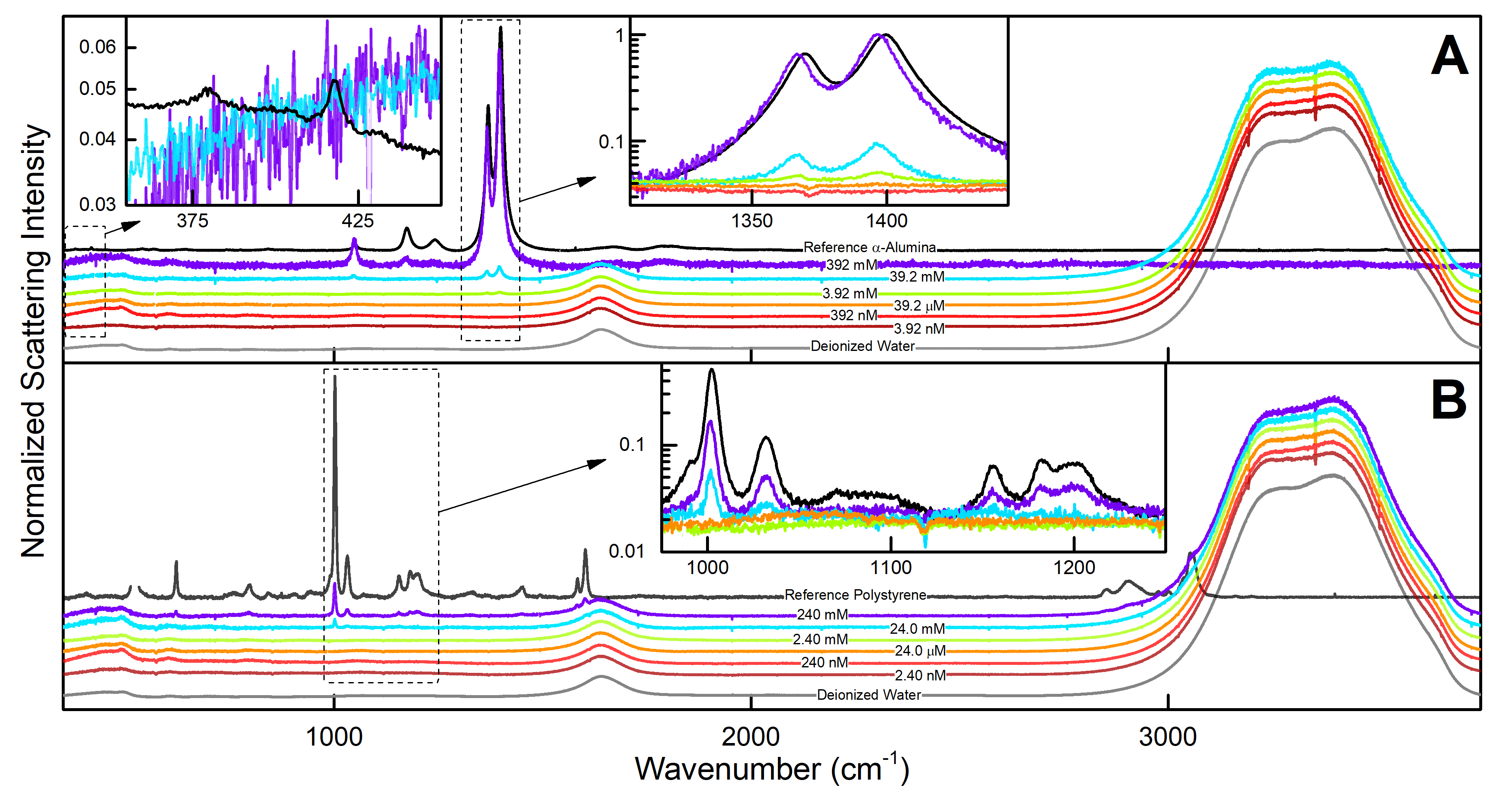}
		\caption{Limit of detection for $\alpha$-Alumina and Polystyrene (PS) nanoparticles when measured in Hollow-Core Photonic Fibers. \textbf{A}: Below 3.92 mM, the luminescent ruby lines at 1367.4 cm$^{-1}$ and 1397.4 cm$^{-1}$ are absent. At all concentrations, the Raman modes at 375 cm$^{-1}$ and 417 cm$^{-1}$ cannot be resolved with confidence. \textbf{B}: At 2.4 mM, the prominent peak at 1001 cm$^{-1}$ is no longer visible and the spectrum appears largely identical to pure water. Spectra in the main plot are manually offset for clarity. Insets: Log-scale detail reveals the limit of detection in the vicinity of characteristic Raman modes.} 
		\label{fig:Prefilm_Limit}
	\end{figure*}
	
	Nanoparticles of $\alpha$-phase Al$_{2}$O$_{3}$ in \gls{ac:diw} provide an example of inorganic Raman enhancement. $\alpha$-Al$_{2}$O$_{3}$ nanoparticles have a wide range of uses in modern technology and drug development, such as acting as an antigen carrier in therapeutic cancer vaccination \cite{Li2011}. \Gls{ac:ps} nanoparticles are utilized to further assess the performance for organic compounds. \Gls{ac:ps} nanoparticles have various applications including but not limited to their use in cancer treatment, facilitating the transport of anti-tumour drugs into cancerous cells \cite{Cabeza2015}.
	
	Spontaneous Raman spectra were collected for several concentrations diluted in pure water to a minimum tested sensitivity of 3.92 nM $\alpha$-Al$_{2}$O$_{3}$ and 2.40 nM \gls{ac:ps}, to show the current limitation using the spontaneous \gls{ac:pcf} technique. These spectra are provided in \textbf{Figure \ref{fig:Prefilm_Limit}}. In the same Figure we show reference spectra for $\alpha$-Al$_{2}$O$_{3}$ and \gls{ac:ps}, obtained by drying the stock nanoparticle dispersion on a silicon wafer and characterizing the resultant film. Signal to noise ratio of these pre-film spectra is limited: this is deliberate to ensure their entire spectrum is captured before there is sufficient time for a film to form. An investigation into the time scales required for thin-film formation is provided in Supporting Information, where we show that the spectrum of a 3.92 mM dispersion begins to show signs of thin-film formation after 1 minute of laser-assisted thermophoresis. It is this thermophoresis effect in conjunction with evaporation and capillary effects at the fibre core walls combined that lead to the formation of the films. All spectra in Figure \ref{fig:Prefilm_Limit} have had baselines removed and are normalized to the height of the most intense peak in the spectrum. Full details of measurement conditions and mode assignment are provided in Methods \cite{Krishnan1947,Porto1967,Liang1958, Cornell1968,Anema2010,Cross1937, Walrafen1964,Carey1998,Choe2016}.
	
	The scattering intensity of Raman modes belonging to each nanoparticle is reduced with decreasing concentration. Figure \ref{fig:Prefilm_Limit}A demonstrates the \gls{ac:lod} for $\alpha$-Alumina. The characteristic $\alpha$-Al$_{2}$O$_{3}$ Raman modes (375 cm$^{-1}$ and 417 cm$^{-1}$) are not visible at any concentration. 
	
	Two modes appear at 1367.4 cm$^{-1}$ and 1397.4 cm$^{-1}$ in the most concentrated samples but are absent below 3.92 mM; these are not Raman scattering signals, but instead correspond to the luminescence of the crystalline ruby structure at 6927 \r{A} and 6942 \r{A} when excited by the 632.8 nm source \cite{Krishnan1947}. At and below the $\mu$M regime, spectra appear identical to \gls{ac:diw}. As the photoluminescent modes do not scale with concentration in a similar manner to Raman, they cannot be used as an indicator for \gls{ac:lod} and thus we must conclude that \gls{ac:lod} $>$ 392 mM. Figure \ref{fig:Prefilm_Limit}B demonstrates the \gls{ac:lod} for \gls{ac:ps}. The characteristic peaks at 1001 cm$^{-1}$ and 1030 cm$^{-1}$ are only visible down to 24 mM. Therefore we conclude the limit of detection for \gls{ac:ps} nanospheres using spontaneous Raman in \gls{ac:pcf} lies in the interval 2.4 mM $<$ \gls{ac:lod} $<$ 24 mM. 
	
	These measurements demonstrate that the sensitivity of the spontaneous Raman signal obtained via optofluidic \gls{ac:pcf}-enhancement does not extend to the micromolar ($\mu$M) regime for these samples.

\section{Coffee-Stain Effect in Photonic Crystal Fibers} % word count 228 + 197 = 425

	The technique described here is developed in response to sensitivity limitations of spontaneous Raman spectroscopy in \glspl{ac:pcf}, where the sample and laser are co-located throughout the full length of the fiber. The laser beam which induces Raman scattering is used for a simultaneous second purpose: promoting efficient thermophoresis in the \gls{ac:pcf} core, where this targeted delivery of energy prompts the nanoparticles in solution to aggregate and form a thin film atop the fiber.
	
	The phenomenon which governs thin-film formation is known as the coffee-stain effect. This effect, which leaves a residual stain around the perimeter of a dried liquid droplet, may be exploited for novel fabrication techniques such as creating dips and rims in colloidal films, or for passive analyte pre-concentration of biological fluids \cite{Deegan1997,Filik2008,Parneix2010}. Conditions for the effect to occur require that the outer perimeter is pinned in place. In this study capillarity is used to transport sample from the reservoir which sits below the \gls{ac:pcf} up to the top facet, and simultaneously ensures that the conditions are met for the coffee-stain effect. It must be highlighted that coffee stain affects on a surface such as Si for example of the same sample type and similar concentration would not allow one to probe the precipitated film to obtain Raman spectra such as those obtained in the fiber core. 
	
	In the context of this experiment, the \gls{ac:pcf} facet is efficiently heated by the Raman excitation laser, which promotes thermophoresis of the sample \cite{Piazza2008}. In the absence of a pinned edge, the outer perimeter would evaporate uniformly and that would only induce evaporation induced effects which require days to achieve similar films for samples with similar concentrations. With a pinned edge, however, fluid which evaporates from the perimeter must be replenished, and a resulting flux transports fluid outwards from the air-liquid boundary at the center of the fiber core \cite{Deegan2000}. This is in contrast to other experiments we carried out to induce spontaneous evaporation in the lab, which took days for the film to form and required orders of magnitude more liquid volume to obtain similar Raman spectra. The time and volume required in the case of the samples discussed here emphasise the dominant role that is played by the process of thermophoresis.
	
	\begin{figure}[t]
		\centering
		\includegraphics[width=0.5\linewidth]{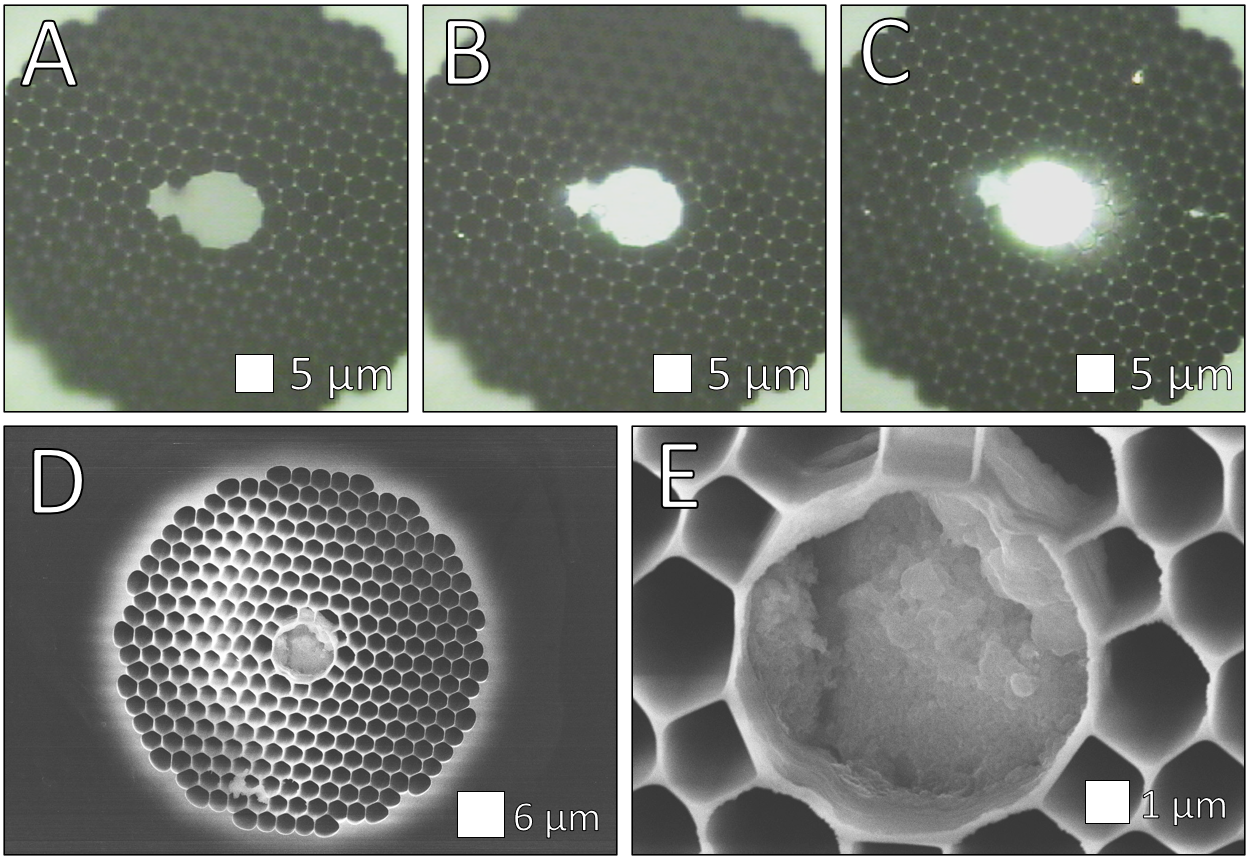}
		\caption{\textbf{A, B, C}: Microscope images showing the progression of thin-film formation. \textbf{A}: The filled core of a \acrfull{ac:pcf} before thermophoresis. \textbf{B}: \gls{ac:pcf} facet after 48 hours of laser-assisted thermophoresis. \textbf{C}: Following thermophoresis, the \gls{ac:pcf} core is removed from the sample reservoir. Nanoparticle thin-film remains atop the fiber core and strongly reflects white light from the microscope. \textbf{D, E}: A scanning electron microscope image of a thin film forming on the core of a \gls{ac:pcf}. The outer rim is the thickest part of the structure, being the initial site of film formation.}
		\label{fig:film}
	\end{figure}
	
	As laser-induced thermophoresis progresses, nanoparticles are transported towards the outer edge they aggregate \cite{Piazza2008}, assisted by capillary and evaporation effects . A white-light microscope image of the waveguide facet taken before and after film formation is shown in \textbf{Figure \ref{fig:film}} (panels A to C), where greater reflection in the core is used as visual confirmation of the film. The thin film is grown inwards from the outside edge. This behaviour is observed in \gls{ac:sem} images of the fiber core post-measurement, seen in Figure \ref{fig:film} (panels D and E), showing increased nanoparticle deposition along the core's outer rim.  
	
	The type of film generated during nanoparticle deposition is influenced by two factors: $C$, speed of the deposition front, and $E$, evaporation rate \cite{Kaplan2015}. When $C \ll E$, the meniscus in the center of a droplet may contact the substrate and a single outer ring or series of rings results. Conversely, when $C \gg E$, deposition outpaces evaporation and a uniform thin film is generated. \Gls{ac:sem} images of our thin films appear to indicate that we are in between these regimes. Exact modelling is beyond the scope of this work; one may refer to Kaplan \textit{et al.} for further information on thin-film modeling \cite{Kaplan2015}. 

\section{Results} % word count 51 + 288 + 325 = 664

	Both Alumina and \gls{ac:ps} nanoparticles demonstrate mM-level or greater \gls{ac:lod} in Figure \ref{fig:Prefilm_Limit} using spontaneous \gls{ac:pcf} Raman. Here we demonstrate that the proposed thermophoresis enhancement principle can characterize these same nanoparticles with an enhanced sensitivity of at least 6 orders of magnitude over optofluidic-assisted spontaneous Raman.
	
	\begin{figure*}[t]
		\centering
		\includegraphics[width=0.9\linewidth]{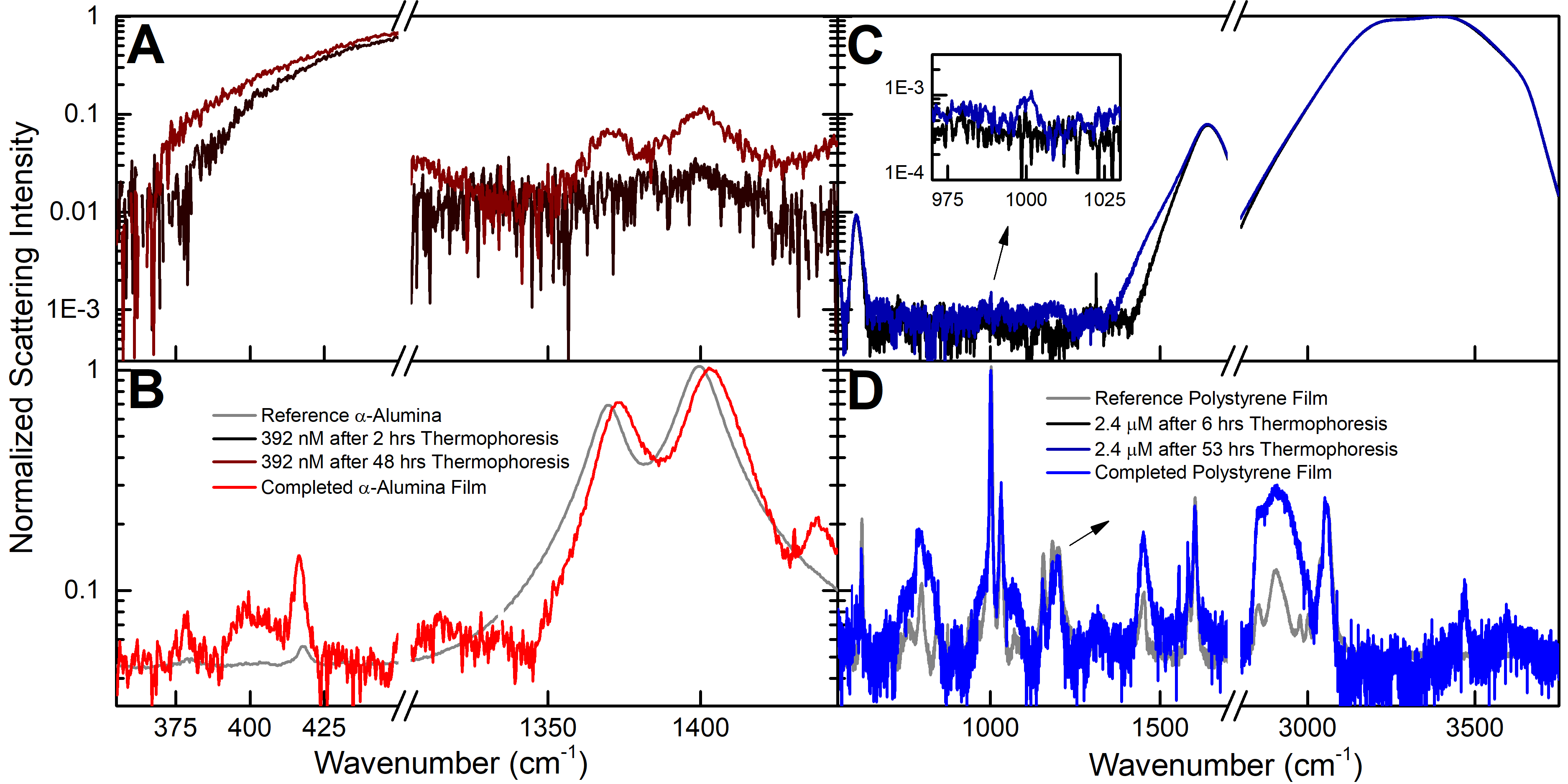}
		\caption{Confirmation of thin-film formation atop a \acrfull{ac:pcf}. \textbf{A}: A solution of $\alpha$-Al$_{2}$O$_{3}$ nanoparticles tested at the nanomolar (392 nM) level. While the core of the \gls{ac:pcf} is filled with sample, no Raman modes are present; luminescent $\alpha$-Al$_{2}$O$_{3}$ modes (circa 1400 cm$^{-1}$) appear after 48 hours of thermophoresis. \textbf{B}: $\alpha$-Al$_{2}$O$_{3}$ Raman modes are revealed after thermophoresis is complete and the fiber core is drained, consistent with the reference spectrum. \textbf{C}: A solution of \acrfull{ac:ps} nanoparticles tested at the micromolar (2.40 $\mu$M) level. The prominent Raman mode at 1001 cm$^{-1}$ is revealed after 53 hours of thermophoresis (inset). \textbf{D}: \Gls{ac:ps} Raman modes are revealed after thermophoresis is complete and the fiber core is drained, consistent with the reference spectrum. All spectra are normalized to the height of the most intense feature in the Raman spectrum.} 
		\label{fig:thermoFilms}
	\end{figure*}
	
\subsection{Alumina Nanoparticles} 

	To demonstrate the formation of an $\alpha$-Al$_{2}$O$_{3}$ thin film via the proposed thermophoresis technique, we present results using a 0.392 nM solution. As evidenced in Figure \ref{fig:Prefilm_Limit}A, this concentration falls below the \gls{ac:lod} for spontaneous Raman in \glspl{ac:pcf}. No Alumina modes, whether Raman or luminescent, were discernible at this concentration when tested using standard \gls{ac:pcf} optofluidic enhancement.
	
	Results from application of the proposed thermophoresis technique are illustrated in \textbf{Figure \ref{fig:thermoFilms}}A. Table \ref{table:alumina} details relevant spectroscopic modes of Alumina found in literature \cite{Krishnan1947,Porto1967}. At the onset of measurements (Figure \ref{fig:thermoFilms}A, black spectrum), no Alumina Raman modes are present and the spectrum is dominated by a strong silica response (from the \gls{ac:pcf}) in the region $<$ 500 cm$^{-1}$. Luminescent modes in the vicinity of 1400 cm$^{-1}$ begin to become apparent after 48 hours of thermophoresis (Figure \ref{fig:thermoFilms}A, dark red spectrum). The presence of these luminescent modes is taken as an indication that the film has successfully formed and the reservoir of fluid can be removed from below the \gls{ac:pcf}, allowing the sample to drain from the fiber core. Once drained, the film remains a permanent fixture atop the fiber and a spontaneous Raman spectrum (Figure \ref{fig:thermoFilms}A, bright red spectrum) reveals the characteristic Raman modes. We can now clearly see two modes at 375 cm$^{-1}$ and 417 cm$^{-1}$, consistent with literature as $E_{g}$ and $A_{1g}$ respectively, as well as the two luminescent modes at 1371 cm$^{-1}$ and 1401 cm$^{-1}$ \cite{Krishnan1947,Porto1967}. In addition to literature, these Raman modes are consistent with the reference spectrum of $\alpha$-Al$_{2}$O$_{3}$ (Figure \ref{fig:thermoFilms}A, grey spectrum).
	
\subsection{PS Nanospheres}

	We wish to demonstrate that the behaviour which governs thin-film formation is not unique to Alumina nanoparticles. Thus we have selected the 2.40 $\mu$M dilution of \gls{ac:ps}, a concentration established to fall below the \gls{ac:lod} for spontaneous Raman in \glspl{ac:pcf} in Figure \ref{fig:Prefilm_Limit}B, to demonstrate increased sensitivity and breadth of applicability of the proposed thermophoresis technique. 
	
	Spontaneous Raman spectra captured during and following the proposed film-formation technique are presented in Figure \ref{fig:thermoFilms}B. Raman mode assignment in polystyrene has historically suffered some contradiction from one publication to the next, so we present here the expected and observed positions of \gls{ac:ps} modes \cite{Anema2010}. We assign the most prominent Raman modes observed, listed in Table \ref{table:psl}, where the first column denotes the band assignment and symmetry, consistent with polystyrene band positions of ordinary Raman by Anema \textit{et al.} \cite{Anema2010}.
	
	At the onset of measurements (Figure \ref{fig:thermoFilms}B, black spectrum), no \gls{ac:ps} Raman modes are present and the spectrum is dominated by modes belonging to \gls{ac:diw}. After 53 hours of thermophoresis (Figure \ref{fig:thermoFilms}B, dark blue spectrum), the prominent Raman mode at 1001 cm$^{-1}$ begins to become apparent, indicating successful film formation. Once the sample is allowed to drain from the \gls{ac:pcf} core, a thin film remains atop the fiber which is confirmed to be \gls{ac:ps} via its spontaneous Raman signal (Figure \ref{fig:thermoFilms}B, bright blue spectrum). The most prominent modes are documented in Table \ref{table:psl} and closely agree with literature for \gls{ac:ps}, primarily coinciding with in- and out-of plane bend and stretch modes of the benzene ring \cite{Liang1958}. In addition to literature, these Raman modes are consistent with the reference spectrum of \gls{ac:ps} (Figure \ref{fig:thermoFilms}B, grey spectrum). With multi-mode agreement between the sample and \gls{ac:ps} literature, we can conclude that detection of \gls{ac:ps} is successful at a micromolar concentration \cite{Liang1958,Cornell1968,Anema2010}.

\section{Discussion And Conclusion} % word count 462
	
	The major distinction between this paper's method and that in Mak \textit{et al.} lies in the deliberate introduction of laser-induced thermophoresis at the laser-facing \gls{ac:pcf} facet \cite{Mak2011}. This process is depicted in Figure \ref{fig:setup}. During prolonged thermophoresis, solute nanoparticles at the fluid surface experience an evaporative flux due to edge-pinning by the \gls{ac:pcf} core. This flux transports nanoparticles to the permiter of the core, where they aggregate into a rigid thin film \cite{Piazza2008, Deegan1997, Kaplan2015}. Using the laser to promote the coffee-ring effect, rather than external application of heat, we enable both in-situ monitoring of thin-film formation, and the precise delivery of energy to the sample where evaporation near the surface takes place. 
	
	This work demonstrates for the first time that a \gls{ac:pcf} coupled with this new thin-film thermophoresis technique can be used to achieve up to $10^6$ mol/L increase in the sensitivity of trace contaminants in aqueous solution. The method is shown to be effective for both organic and inorganic samples. The coffee-ring effect has been experimentally verified for both aqueous and non-aqueous fluids, and for analyte sizes ranging from the molecular up to 10 $\mu$m \cite{Deegan2000}. We were unable to replicate these results when dispersions of larger (1 $\mu$m) particles are used in the capillary fiber. A thorough investigation of size limitations for thin-film thermophoresis, and solutions to overcome them, is beyond the scope of this work. This limitation does not appear to affect the nanoparticle regime, and we anticipate that thermophoretic thin-film formation is feasible across a range of samples much wider than those presented here. 
	
	The size dependence described above, along with the evidence provided in the results section helps form the picture of how the laser not only simply accelerates evaporation, through heating but also enables rapid film formation with orders of magnitude more film volume per unit time due to another affect that is not evaporation related. These arguments points to the dominant role of   thermophoresis over others including evaporation, capillary effects and others.
	
	Relative to established methods of nanoparticle characterization, thin-film thermophoresis offers a great deal of flexibility. X-Ray Diffraction, for example, is limited to particles under 50 nm in size \cite{Mourdikoudis2018}. This poses an issue for gold nanoparticles: uptake rate into mammalian cells is highly dependent on particle size, including those up to 100 nm, making that technique incompatible \cite{Chithrani2006}. Alternatively, transmission electron microscopy provides more precise determination of particle size, arrangement, and agglomeration. It generates a direct image of nanoparticles, but this does not reveal the chemical identity of the sample, and imaging parameters must be carefully determined in order to prevent sample damage and to be certain the image is not skewed with respect to its orientation, providing misleading information \cite{Pyrz2008,Mourdikoudis2018}. Comparatively, thin-film thermophoresis is simple to perform and provides direct confirmation of the particle identity.
	
	This work sets the stage for exciting new directions. At the most fundamental level, it offers the ability to achieve the same order of magnitude enhancement as conventional \gls{ac:sers}, with more repeatability and less complexity. Potential extensions of this work include a novel hybrid method to form films compatible with \gls{ac:sers}, providing additional sensitivity beyond that demonstrated here. 
	
\section{Experimental Section} % word count 483

\threesubsection{Spectrum Collection}

	In both the benchmark and proposed thermophoresis technique, we use a \gls{ac:pcf} (HC-800, NKT Photonics) as the \gls{ac:pcf}. A beaker of aqueous sample is set on the microscope stage beneath the \gls{ac:pcf}, which has been uniquely prepared in such a way that impedes the flow of liquid from entering the cladding holes and only permits fluid to infiltrate the core. This fashion of preparation ensures that best accuracy is obtained when determining the LOD. This solution is transported via capillary action to the top cross section of the fiber tip, where it is illuminated by the laser, which is used for both Raman measurements and, in the instances of film formation, it is also utilized for thermophoresis. All Raman measurements are performed on an Horiba HR800 Raman Spectrometer using a 632.8 nm laser and a grating of 1800 lines/mm. Samples depicted in Figure \ref{fig:Prefilm_Limit} were captured using a 100$x$  objective with 1 second exposure and averaging of 5 spectra. A short spectrum collection time and optical density filter (1.1 mW incident on sample) are necessary for these samples in order to inhibit thin-film formation via thermophoresis, capturing a spectrum which is truly representative of the sample in its native state. Spectra collected during thermophoresis were excited with full laser power (11.2 mW incident on sample), 2-3 seconds exposure time and averaging up to 75 spectra, in order to deposit as much energy as possible to the sample without saturating the detector.

\threesubsection{Nanoparticle Raman Mode Assignments}

\threesubsection{Alumina}

	\textbf{Table \ref{table:alumina}} details relevant spectroscopic modes of Alumina found in literature \cite{Krishnan1947,Porto1967}. The modes at 1367.4 cm$^{-1}$ and 1397.4 cm$^{-1}$ are not Raman scattering signals, but instead correspond to the luminescence of the crystalline ruby structure at 6927 \r{A} and 6942 \r{A} when excited by the 632.8 nm source \cite{Krishnan1947}.
	
	\begin{table}[htbp]
		\centering
		\caption{Raman Mode Assignments for Alumina.}
		\label{table:alumina}
		\begin{tabular}{c c}
			\hline 
			Assignment			& Wavenumber (cm$^{-1}$)\\
			\hline 
			$E_{g}$  			& 375\\
			$A_{1g}$ 		 	& 417\\
			Luminescence		& 1367.4\\
			Luminescence		& 1397.4\\
			\hline
		\end{tabular}
	\end{table}

\threesubsection{Polystyrene}

	Raman mode assignment in polystyrene has historically suffered some contradiction from one publication to the next, so we present here the expected and observed positions of \gls{ac:ps} modes \cite{Anema2010}. We assign the most prominent Raman modes observed in the first loading vector obtained from running Principal Component Analysis on the spectra from \gls{ac:ps} at 240 mM (pre-film) and 2.40 $\mu$M (post-film). This vector represents the most dominant common spectral features between the two samples, which are expected to be consistent with \gls{ac:ps} in literature. Raman modes and their assignments are listed in \textbf{Table \ref{table:psl}}, where the first column denotes the band assignment (labelled using Wilson notation) and symmetry, consistent with polystyrene band positions of ordinary Raman by Anema \textit{et al.} \cite{Anema2010}.
	
	% Notation reference: https://en.wikipedia.org/wiki/List_of_character_tables_for_chemically_important_3D_point_groups
	\begin{table}[ht]
		\centering
		\caption{Raman Mode Assignments for Polystyrene.}
		\label{table:psl}
		\begin{tabular}{c c c}
			\hline 
			Assignment			& \multicolumn{2}{c}{Wavenumber (cm$^{-1}$)} \\
			\cline{2-3}
								& Expected \cite{Liang1958,Anema2010}	& Observed	\\
			\hline 
			6b $ B_2 $ 			& 623				& 619.7 	\\
			12 $ A_1 $			& 1002				& 1000.6	\\
			18a $ A_1 $			& 1032				& 1030.3	\\
			15 $ B_2 $			& 1156				& 1154.3	\\
			9a $ A_1 $			& 1183				& 1181.1	\\
			13 $ A_1 $			& 1196				& 1198.5	\\
			8b $ B_2 $			& 1583				& 1581.9 	\\
			8a $ A_1 $			& 1600				& 1601.1	\\
			7a $ A_1 $			& 3056				& 3051.1	\\	
			\hline
		\end{tabular}
	\end{table}	
	
\threesubsection{Determination of the limit of detection}

In order to determine the LOD, Raman spectra with the same conditions including, power, time, objective lens among others has been collected for samples with progressively reduced concentrations. The spectra were then analysed, where peak search using the algorithms discussed in the paper were utilized to quantify the relevant Raman modes for every sample under test. The limit of detection is determined for a given material when no modes could be identified in the spectrum below a certain concentration. It is important that we identified a given  when we are able to find more than one mode of its vibrational spectrum, and not just one mode. While this technique is not absolute, as it is bound by the parameters used such as power and time, it is indicative of the mechanism discussed in this work. As we can increase the detection time or the utilized power significantly, the times used in this work, and first report, highlight the utility of Thermophresis-based approach to form films and increase the tenable LOD of spontaneous Raman in liquids for measurements with practical time durations. More absolute limits can and will be  examined in dedicated work.

\medskip
\textbf{Conflict of Interest} \par
The authors declare no conflicts of interest.

\medskip
\textbf{Supporting Information} \par %Please delete the Suppporting Information statement if it is not applicable. Please supply Supporting Information in another file. Supporting information should not be provided in .tex format
Supporting Information is available from the Wiley Online Library.

% References
\medskip

% Use the following code if you wish to generate your bibliography with BibTeX;
% replace the string "MSP-template" below with the name(s) of
% the BibTeX data base(s) you want to use.
% The resulting bibliography-output (the content of the .bbl file)
% must be pasted back into this file before submission.
% Please also include your BibTeX data base file(s) in your submission
% so that we can re-run BibTeX if necessary.
%
\bibliographystyle{MSP}
\bibliography{BasilBib}

\begin{thebibliography}{10}
\providecommand{\url}[1]{\texttt{#1}}
\providecommand{\urlprefix}{URL }

\bibitem{Cabeza2015}
L.~Cabeza, V.~Cano-Cort{\'{e}}s, M.~J. Rodr{\'{i}}guez, C.~V{\'{e}}lez,
  C.~Melguizo, R.~M. S{\'{a}}nchez-Mart{\'{i}}n, J.~Prados,
\newblock \emph{Journal of Nanoparticle Research} \textbf{2015}, \emph{17}, 1.

\bibitem{Burklew2012}
C.~E. Burklew, J.~Ashlock, W.~B. Winfrey, B.~Zhang,
\newblock \emph{PLoS ONE} \textbf{2012}, \emph{7}, 5 34783.

\bibitem{Hood2004}
E.~Hood,
\newblock \emph{Environmental Health Perspectives} \textbf{2004}, \emph{112},
  13 A740.

\bibitem{Mourdikoudis2018}
S.~Mourdikoudis, R.~M. Pallares, N.~T.~K. Thanh,
\newblock \emph{Nanoscale} \textbf{2018}, \emph{10}, 27 12871.

\bibitem{Vandenabeele2007}
S.~N. Bertram, B.~Spokoyny, D.~Franks, J.~R. Caram, J.~J. Yoo, M.~R., M.~E.
  Grein, M.~G. Bawendi,
\newblock \emph{ACSnano} \textbf{2019}, \emph{13}, 3 1024.

\bibitem{DeBeer2011}
W.~Lu, H.~S. Wu, X.~G., L.~Wang, Q.~Zhou, Z.~B., A.~Zhong, Y.~Wang,
\newblock \emph{International Journal of Pharmaceutics} \textbf{2071},
  \emph{5}, 1-2 1700594.

\bibitem{Evans2008}
E.~Storey, D.~Wu, A.~S. Helmy,
\newblock \emph{Annual Review of Analytical Chemistry} \textbf{2021},
  \emph{39}, 17 5634.

\bibitem{Tang2018}
E.~Storey, A.~S. Helmy,
\newblock \emph{Journal of Raman Spectroscopy} \textbf{2019}, \emph{50}, 8 958.

\bibitem{Gouadec2007}
G.~Gouadec, P.~Colomban,
\newblock \emph{Progress in Crystal Growth and Characterization of Materials}
  \textbf{2007}, \emph{53}, 1 1.

\bibitem{Fan2020}
M.~Fan, G.~F. Andrade, A.~G. Brolo,
\newblock \emph{Analytica Chimica Acta} \textbf{2020}, \emph{1097} 1.

\bibitem{Lafuente2020}
M.~Lafuente, D.~Sanz, M.~Urbiztondo, J.~Santamar{\'{i}}a, M.~P. Pina,
  R.~Mallada,
\newblock \emph{Journal of Hazardous Materials} \textbf{2020}, \emph{384}
  121279.

\bibitem{Mak2013}
J.~S. Mak, S.~A. Rutledge, R.~M. Abu-Ghazalah, F.~Eftekhari, J.~Irizar, N.~C.
  Tam, G.~Zheng, A.~S. Helmy,
\newblock \emph{Progress in Quantum Electronics} \textbf{2013}, \emph{37}, 1 1.

\bibitem{Mak2011}
J.~S.~W. Mak, A.~A. Farah, F.~Chen, A.~S. Helmy,
\newblock \emph{ACS Nano} \textbf{2011}, \emph{5}, 5 3823.

\bibitem{Li2011}
H.~Li, Y.~Li, J.~Jiao, H.-M. Hu,
\newblock \emph{Nature Nanotechnology} \textbf{2011}, \emph{6}, 10 645.

\bibitem{Krishnan1947}
R.~S. Krishnan,
\newblock {Raman spectrum of alumina and the luminescence and absorption
  spectra of ruby [14]}, \textbf{1947}.

\bibitem{Porto1967}
S.~P. Porto, R.~S. Krishnan,
\newblock \emph{The Journal of Chemical Physics} \textbf{1967}, \emph{47}, 3
  1009.

\bibitem{Liang1958}
C.~Y. Liang, S.~Krimm,
\newblock \emph{Journal of Polymer Science} \textbf{1958}, \emph{27}, 115 241.

\bibitem{Cornell1968}
S.~W. Cornell, J.~L. Koenig,
\newblock \emph{Journal of Applied Physics} \textbf{1968}, \emph{39}, 11 4883.

\bibitem{Anema2010}
J.~R. Anema, A.~G. Brolo, A.~Felten, C.~Bittencourt,
\newblock \emph{Journal of Raman Spectroscopy} \textbf{2010}, \emph{41}, 7 745.

\bibitem{Cross1937}
P.~C. Cross, J.~Burnham, P.~A. Leighton,
\newblock \emph{Journal of the American Chemical Society} \textbf{1937},
  \emph{59}, 6 1134.

\bibitem{Walrafen1964}
G.~E. Walrafen,
\newblock \emph{The Journal of Chemical Physics} \textbf{1964}, \emph{40}, 11
  3249.

\bibitem{Carey1998}
D.~M. Carey, G.~M. Korenowski,
\newblock \emph{The Journal of Chemical Physics} \textbf{1998}, \emph{108}, 7
  2669.

\bibitem{Choe2016}
C.~Choe, J.~Lademann, M.~E. Darvin,
\newblock \emph{The Analyst} \textbf{2016}, \emph{141}, 22 6329.

\bibitem{Deegan1997}
R.~D. Deegan, O.~Bakajin, T.~F. Dupont, G.~Huber, S.~R. Nagel, T.~A. Witten,
\newblock \emph{Nature} \textbf{1997}, \emph{389}, 6653 827.

\bibitem{Filik2008}
J.~Filik, N.~Stone,
\newblock \emph{Analytica Chimica Acta} \textbf{2008}, \emph{616}, 2 177.

\bibitem{Parneix2010}
C.~Parneix, P.~Vandoolaeghe, V.~S. Nikolayev, D.~Qu{\'{e}}r{\'{e}}, J.~Li,
  B.~Cabane,
\newblock \emph{Physical Review Letters} \textbf{2010}, \emph{105}, 26 266103.

\bibitem{Piazza2008}
R.~Piazza, A.~Parola,
\newblock \emph{Journal of Physics: Condensed Matter} \textbf{2008}, \emph{20},
  15 153102.

\bibitem{Deegan2000}
R.~D. Deegan, O.~Bakajin, T.~F. Dupont, G.~Huber, S.~R. Nagel, T.~A. Witten,
\newblock \emph{Physical Review E} \textbf{2000}, \emph{62}, 1 756.

\bibitem{Kaplan2015}
C.~N. Kaplan, L.~Mahadevan,
\newblock \emph{Journal of Fluid Mechanics} \textbf{2015}, \emph{781} R2.

\bibitem{Chithrani2006}
B.~D. Chithrani, A.~A. Ghazani, W.~C. Chan,
\newblock \emph{Nano Letters} \textbf{2006}, \emph{6}, 4 662.

\bibitem{Pyrz2008}
W.~D. Pyrz, D.~J. Buttrey,
\newblock \emph{Langmuir} \textbf{2008}, \emph{24}, 20 11350.

\end{thebibliography}

% Table of contents entry should be 50 - 60 words long
% Image should be 55 mm broad and 50 mm high or 110 mm broad and 20 mm high

\begin{figure}[h]
\textbf{Table of Contents}\\
%\medskip
  \includegraphics[width=55mm]{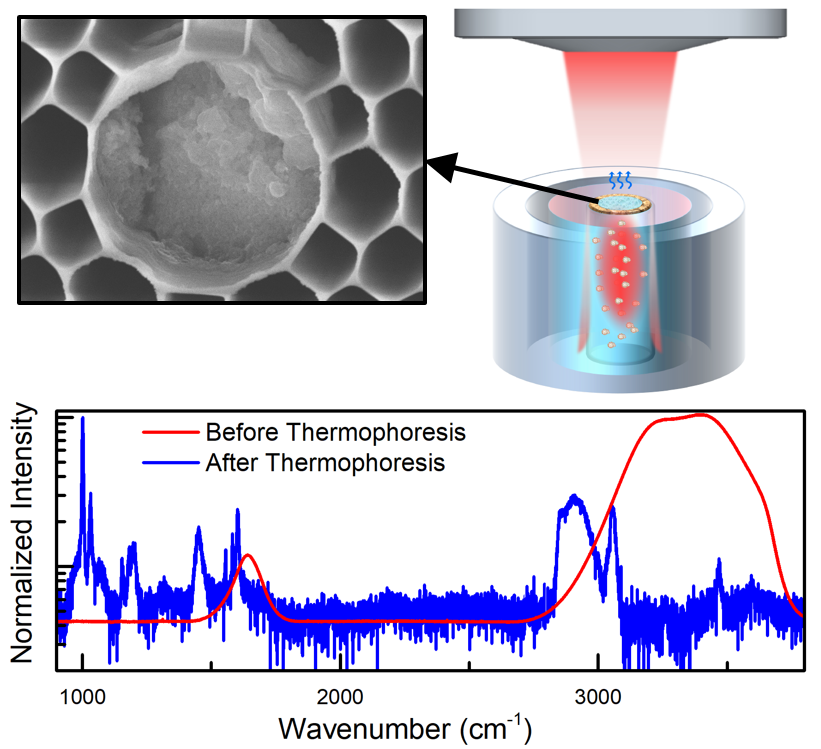}
  \medskip
  \caption*{Enhanced spontaneous Raman sensitivity through the formation of a thin film via thermophoresis atop a Hollow-Core Photonic Crystal Fiber is reported for the first time. Sensitivity of detection is increased by more than 6 orders of magnitude. Detection of two nanoparticles, Alumina and Polystyrene, is demonstrated down to 392 nM without the use of Surface-Enhanced Raman Spectroscopy.} % word count = 58
\end{figure}

\end{document}